\documentstyle[epsfig,prb,aps,multicol]{revtex}

\begin{document}

\draft
\title{Enhancement of the magnetic anisotropy of nanometer-sized Co
clusters: influence of the surface and of the inter-particle
interactions}

\author{F. Luis$^{1}$, J.M. Torres$^{1}$, L.M. Garc\'{\i}a$^{1}$, J. Bartolom\'e$^{1}$,
J. Stankiewicz$^{1}$,  F. Petroff$^{2}$, F. Fettar$^{2 \dag}$,
J.-L- Maurice$^{2}$, and A. Vaur${\em \grave{e}}$s$^{2}$}

\address{$^1$ Instituto de Ciencia de Materiales de Arag\'on,
CSIC-Universidad de Zaragoza, 50009 Zaragoza, Spain\\ $^2$Unit\'e
Mixte de Physique CNRS/THALES, UMR 137, Domaine de Corbeville,
91404 Orsay Cedex, France\\ $^{\dag}$Present address: Laboratoire
de Nanostructures et Magn\'etisme, DRFML/SP2M, CEA, 38054 Grenoble
Cedex 9, France}

\date{\today}

\maketitle

\begin{abstract}
We study the magnetic properties of spherical Co clusters with
diameters between $0.8$ nm and $5.4$ nm ($25$ to $7500$ atoms)
prepared by sequential sputtering of Co and Al$_{2}$O$_{3}$. The
particle size distribution has been determined from the
equilibrium susceptibility and magnetization data and it is
compared to previous structural characterizations. The
distribution of activation energies was independently obtained
from a scaling plot of the ac susceptibility. Combining these two
distributions we have accurately determined the effective
anisotropy constant $K_{eff}$. We find that $K_{eff}$ is enhanced
with respect to the bulk value and that it is dominated by a
strong anisotropy induced at the surface of the clusters.
Interactions between the magnetic moments of adjacent layers are
shown to increase the effective activation energy barrier for the
reversal of the magnetic moments. Finally, this reversal is shown
to proceed classically down to the lowest temperature investigated
($1.8$ K).
\end{abstract}

PACS:75.50.Tt,75.70.-i,75.40.Gb,75.30.Pd

\newpage

\section{INTRODUCTION}
Single domain magnetic particles are attractive for applications
in data storage. Their properties differ from those of the bulk
magnets\cite{Dormann97} because, as the size of the particles
decreases, an increasing fraction of the total magnetic atoms lies
at the surface. The electronic and magnetic  structure of these
atoms can be modified by the smaller number of neighbors as
compared to the bulk\cite{Hendricksen92,Reddy93,Pastor95} and/or
by the interaction with the surrounding atoms of the matrix where
the particles are dispersed. For example, it was shown by Van
Leeuwen et al. that the bonding of CO at the surface of Ni
clusters induces quenching of the magnetic moments of those atoms
located at the surface.\cite{Leeuwen94} In some other cases, the
surface layer is oxidized and shows antiferro or spin-glass like
arrangement of the magnetic moments, which also leads to a smaller
net magnetic moment of the particle.\cite{Kodama96} By contrast,
measuring "bare" particles of Fe, Co and Ni produced in beams, de
Heer and coworkers\cite{Billas94} found that the net magnetic
moment per atom increases as the size of the cluster decreases,
approaching the limiting value for a free atom. In addition, the
net anisotropy of the particle exceeds the bulk
value.\cite{Bodker94,Chen95} This excess was recently correlated
to the augmentation of the orbital magnetic moment of the
peripheral atoms.\cite{Durr99,Edmons99}

Magnetic nanoparticles are also
good candidates for the study of quantum effects in intermediate
scales between the microscopic and the macroscopic "classical"
world.\cite{Gunther95,Chudnovsky98} In real systems however,
we usually deal with macroscopic ensembles of particles with different sizes
and shapes. The average magnetic properties of these systems
come from intra-particle as well as interparticle phenomena, which are usually difficult to
disentangle. Therefore, in this field of research it is desirable to obtain
systems in which each of the parameters, such as the average particle
size, the particle size distribution, the crystalline structure, and the
spatial arrangement of the particles, can be varied independently
of each other.

We believe that the work reported here is a step forward in this
direction. We present the magnetic characterization of a new type
of systems of Co nanoparticles, embedded in an amorphous matrix of
Al$_{2}$O$_{3}$, prepared by sequential deposition of both
materials.\cite{Maurice99,Briatico99,Babonneau00} By varying the
deposition time, the diameter of the aggregates can be controlled
between below $1$ nm and $7$ nm. An important advantage of this
preparation method is that it gives a rather homogeneous
dispersion of the particles in the matrix. It was also found that,
for a range of thicknesses,  a relatively ordered disposition of
the particles is obtained, in which they are arranged in layers
separated from the adjacent ones by a controllable
distance.\cite{Babonneau00} The paper is organized as follows. In
the first two sections we briefly describe the method employed to
prepare the samples and their physical characterization. Then, we
present our experimental results. Using the data obtained from ac
magnetic susceptibility, zero-field cooled (ZFC) and field-cooled
(FC) magnetization measurements, and isotherms of magnetization as
a function of the field, we have determined the particle size
distribution in samples which have been prepared with different Co
deposition times. We compare these results with available data
from a previous structural characterization. This important
information is then used to determine accurately the effective
anisotropy constant and its variation with the size of the
particles. We have also been able to separate surface anisotropy
effects from the effect of the dipole-dipole interaction between
the magnetic moments of the particles. The last section is left
for the conclusions.

\section{MORPHOLOGY AND STRUCTURE OF THE SAMPLES}
\label{SSamples}

Details of the sample preparation and of its structural
characterization have already been reported
elsewhere.\cite{Maurice99,Briatico99,Babonneau00} The Co
aggregates were prepared by sputter deposition of Co atoms on a
smooth alumina surface. The amount of deposited Co is given here
by the nominal thickness $t_{Co}$ that the deposits would have if
they were homogeneous. This amount was measured by using
energy-dispersive X-ray spectroscopy in the transmission electron
microscope, and found to be within less than $5$\% of the planned
dose in all cases. Clusters are formed below the percolation limit
which appears to occur at $t_{Co}=2$ nm. On top of each Co layer a
new alumina layer of about $3$ nm was deposited. Oxidized Si was
used as a substrate. A given sample is usually made by piling up a
number $N$ ($1$ to $100$ for the samples studied here) of these
layers.\cite{Babonneau00} The deposition rates of both Co and
alumina were respectively $0.114 \AA$ s$^{-1}$ and $0.43 \AA$
s$^{-1}$. It was found that the amount of Co deposited on the
surface is larger than the Co mass which forms clusters visible by
transmission electron microscopy (TEM). The relative difference
between these quantities increases as $t_{Co}$ decreases.
Therefore, we have in our samples non-aggregated atoms or very
small clusters, which contribute to the magnetic signal of the
samples, in addition to Co aggregates. One of the difficulties of
the interpretation of the magnetic data is to separate these two
contributions.

Because of the upper alumina layer the aggregates show no trace of
oxide even after exposure to air. The lack of oxide was checked by
electron energy-loss spectroscopy (EELS), X-ray photoelectron
spectroscopy and X-ray absorption spectroscopy. The atomic
structure of the clusters was determined by extended X-ray
absorption fine structure (EXAFS) spectroscopy and high resolution
TEM.\cite{Maurice99} It was found that the particles bond poorly
to the alumina matrix, and that the Co crystallizes in the fcc
phase for $t_{Co} < 1$ nm. The presence of a fcc phase in place of
the hcp phase which is the stable phase for bulk Co is not
uncommon for small particles. It was theoretically
predicted\cite{Kitakami97} and also found
experimentally\cite{Granquist76} that the fcc phase becomes more
stable below some diameter which depends on the matrix.

The morphology, size and spatial distribution of the aggregates
were also studied using the TEM data. The aggregates are of
approximately spherical shape (at least for $t_{Co} < 1$ nm). The
average diameter $\langle D \rangle$ of the particles increases
linearly with $t_{Co}$. We give in Table I a list of the important
parameters obtained from these experiments for all samples
studied. Finally, the TEM pictures reveal a quasi-ordered
arrangement of the Co clusters\cite {Babonneau00} that is induced
by the topology of the layers: the clusters of a layer nucleate
preferentially in the hollows left by the previous layer. In each
layer, the average distance between the borders of adjacent
clusters is of order $2$ nm and approximately independent of
$t_{Co}$.

\section{EXPERIMENTAL DETAILS}
\label{SExp}

The magnetic measurements were performed using a commercial SQUID
magnetometer. The temperature range of the measurements was $1.8$
K $<T< 320$ K and magnetic fields up to $5$ T could be applied by
means of a superconducting magnet. The ac susceptibility was
measured by applying a small ac field ($4.5$ Oe) to the sample and
using the ac detection option of the same magnetometer. The
frequency $\omega/2 \pi$ of the ac magnetic field can be varied
continuously between $0.01$ Hz and $1.5$ kHz. The samples had a
rather large diamagnetic signal arising from the silicon
substrate. This contribution was estimated independently by
measuring a bare substrate and found to be linear in field and
independent of the temperature. It was subsequently subtracted
from all experimental data. Unless indicated otherwise, the data
shown in this paper were measured on samples having more than $20$
Co/Al$_{2}$O$_{3}$ bi-layers in order to maximize their magnetic
signals. We have checked for $t_{Co}=0.3$ nm and $t_{Co}=0.7$ nm
that the variation of the magnetization and of the ac
susceptibility with temperature and magnetic field is rather
insensitive to the precise value of $N$, provided that $N$ is
larger than $10$ layers.

\section{RESULTS AND DISCUSSION}
\label{SRes}

\subsection{Superparamagnetic blocking}
\label{SS.blocking} The magnetic dc susceptibility was measured by
cooling the samples in zero field (ZFC) or in the presence of the
measuring magnetic field (FC). Typical ZFC-FC magnetization curves
are plotted in Fig. 1. At high temperatures the ZFC and FC curves
coincide, indicating that the samples behave as superparamagnets.
In this region, both curves follow the Curie-Weiss law
$C/(T-\theta)$. The value of $C$ increases as $t_{Co}$ increases
(see the inset of Fig. 1), as expected for larger clusters
formation as the deposition time of Co increases. The Curie-Weiss
temperature $\theta$ is nearly zero but for the two samples
containing the largest particles. This is brought about by the
interaction between the particles, which we shall consider in a
separate section below. At lower temperature, the two curves start
to separate. The ZFC curve shows a maximum at a temperature
$T_{B}$ below which the magnetic moments are blocked in fixed
directions. It is well known that the phenomenon of blocking is
related to the magnetic anisotropy of the particles.\cite{Neel49}
The anisotropy favors some particular orientations of the magnetic
moment, two opposite to each other in the simplest case of
uniaxial anisotropy, which are separated by activation energy
barriers $U$. As the temperature decreases, the number of thermal
phonons of energy equal or larger than $U$ decreases, thus leading
to an exponential increase of the time $\tau$ needed to reverse
the magnetic moment of a particle\cite{Neel49,Brown63,Coffey94}
\begin {equation}
\tau=\tau_{0} \exp \left(U/k_{B}T \right) \label{Arrhenius}
\end{equation}
Here $\tau_{0}\approx 10^{-10}-10^{-13}$ s is an inverse
attempt-frequency, which depends on the damping of the magnetic
moment by the phonon or the magnon baths. In this simple picture,
the superparamagnetic blocking takes place when $\tau$ equals the
measurement time of each experimental point $t_{e}$, thus $T_{B}
\simeq \alpha U/k_{B} \ln(t_{e}/\tau_{0})$, where $\alpha$ is a
constant which depends on the width of the particle size
distribution (more details are given  below). We have indeed
observed that $T_{B}$ increases with the Co deposition time, that
is, with the average volume of the aggregates. Therefore, we write
$U = K_{eff} V$, where $K_{eff}$ is an effective anisotropy
constant with contributions from the intrinsic magnetocrystalline
anisotropy of the fcc Co and from other sources, such as stress
induced anisotropy or surface induced anisotropy. The dependence
of $K_{eff}$ on $V$ will be considered below in section
\ref{SS.anisotropy}.

As expected, the blocking of the magnetic moments by the
anisotropy also leads to a maximum in the temperature dependence
of both the real and the imaginary components of the ac magnetic
susceptibility. A typical experimental result is shown in Fig. 2.
The position of the susceptibility peak shifts towards lower
temperatures as the frequency of the ac magnetic field decreases
since $t_{e}$ equals $1/\omega$.

Above $T_{B}$ the magnetization isotherms are fully reversible
because the magnetic moments are in thermal equilibrium. As shown
in Fig. 3, the experimental data measured well above the blocking
temperature of each sample collapse into a single curve when they
are plotted as a function of $H/T$, indicating that the effect of
the anisotropy is weak. Furthermore, pure Langevin curves fit the
experimental data reasonably well, which shows that the size
distributions of all these samples are narrow. Below $T_{B}$, the
magnetization shows hysteresis (see Fig. 4) with both the coercive
field $H_{c}$ and the remanence $M_{r}$ increasing as the
temperature decreases (see Fig. 5).

We plot in Fig. 6 the low-T values of $M_{r}$ and of the
saturation magnetization $M_{s}$ as a function of $t_{Co}$. It is
interesting to note that, for $t_{Co} < 0.7$ nm, the reduced
remanence $m_{r}=M_{r}/M_{s}$ is smaller than the value $1/2$
predicted by the Stoner-Wolhfarth model.\cite{Stoner48} We
attribute the decrease of $m_{r}$ to a paramagnetic contribution,
which adds to that of the blocked particles. A Curie tail shown by
the saturation magnetization at the lowest temperatures (cf Fig.
5) is also related to this extra contribution. The excess
paramagnetism arises likely from single atoms or very small
clusters that are formed in the first stages of the preparation
process and which do not give rise to further
aggregation.\cite{Maurice99,Briatico99,Carrey00} It was found that
the fraction $x_{para}$ of Co which is deposited but is not
detected by TEM increases as $t_{Co}$ decreases. Accordingly,
$m_{r}$ decreases as the amount of deposited Co decreases. On the
other hand, the sample with $t_{Co} = 1$ nm has $m_{r} = 0.71$,
that is, larger than $1/2$, likely because of the predominant
ferromagnetic coupling between particles.

It is also remarkable that the average magnetic moment per atom
for the whole sample, as obtained from $M_{s}$ of Fig. 6, is
smaller than the value for bulk Co ($1.7 \mu_{B}$ per Co atom) for
all samples and that it decreases as $t_{Co}$ and, thus as the
average size of the particles decrease. This dependence is
opposite to that observed for free Co clusters in
beams\cite{Billas94} and also for Co particles of similar size
supported in a solid matrix.\cite{Chen95,Respaud98} In those
experiments, the measured magnetic moment per atom exceeded the
bulk value and it was found to increase as the diameter of the
particles decreases.

The reduced value of $M_{s}$ that we measure could be caused by an
oxide layer at the surface of the particles, which orders
antiferromagnetically. However, as we mentioned before, we did not
find any trace of oxide in EELS measurements. Moreover, it is
known that the exchange interaction between this layer and the
magnetic core of the particles would also induce a net anisotropy
on the latter.\cite{Berkowitz00} This so-called exchange
anisotropy leads to a shift of the hysteresis loops when the
sample is cooled down in the presence of a magnetic field. For
example, Peng and coworkers\cite{Peng99} have recently measured an
exchange bias field as large as $10.2$ kOe for CoO coated Co
clusters having a diameter of $6$ nm and $13$ nm. By contrast, as
we show in Fig. 4, the hysteresis loops measured after cooling the
sample in zero field or in $5$ T from room temperature are nearly
identical, thus with no evidence for an antiferromagnetic order at
the surface layer. Thus, we conclude that most of the particles
are free from oxidation.

It is however still possible that some of the Co atoms, in close
contact with the Al$_{2}$O$_{3}$ matrix, have a weak chemical link
with it. This chemical bonding can reduce the number of unpaired
electrons and then quench the magnetic moment of the metal atom,
as was shown by van Leeuwen and co-workers.\cite{Leeuwen94} From
our data, it is not possible to determine whether the atoms
involved in the reduction of the average magnetic moment are
located at the periphery of the particles or are those atoms which
do not form aggregates, because the relative concentration of both
increases as the average size of the clusters decreases.

Therefore, in what follows, we approach the problem in a different
way. As a starting point of the analysis we consider that the
spheres have the bulk magnetization: $M_{sb} = 1.7 \mu_{B}$ per Co
atom, whereas the missing magnetic moment is exclusively
attributed to the paramagnetic Co fraction. The contribution of
the clusters to the net saturation magnetization of each sample
then equals $(1-x_{para}) M_{sb}$. It was obtained by subtracting
the low-T paramagnetic tail from the total $M_{s}$. In this way,
$x_{para}$ is also estimated. We list the results in Table I. The
low-$T$ paramagnetic magnetization is found to be compatible with
a free spin $1/2$ for all samples, which indicates that the
isolated atoms have in average only one unpaired electron.

In order to fit the magnetic data, it is necessary to know the
fraction $x_{para}$ of paramagnetic atoms and their magnetic
moment.

\subsection{Determination of the particle size distribution}
\label{SS.fv} In this section we will try to determine the
particle size distribution from the equilibrium magnetic
properties of each sample and compare it with the results obtained
by TEM. In Fig. 3 we have plotted the equilibrium magnetization
$M$ of two different samples having $t_{Co}=0.3$ nm and $0.7$ nm,
respectively. We recall that, for a set of magnetic moments $\mu$
without anisotropy, $M(H,T,\mu) = M_{s} L(\mu H/T)$, where $L$
denotes the Langevin function. If the anisotropy energy is taken
into account, there is no analytical expression for $M(H,T,\mu)$
and the shape of the magnetization curve deviates from the pure
Langevin form when $U/k_{B}T$ is large. However, it can still be
evaluated numerically, as was described in Ref. \cite{Hanson93}.
As we said above, the good scaling on a single curve of data
measured at different temperatures indicates therefore that the
influence of the anisotropy is not very important.

For a real sample, we have to average $M(H,T,\mu)$
over the appropriate distribution of particle's sizes. Comparing the calculated
magnetization to the experimental data we shall try next
to get information about this distribution.
In order to directly compare our results with those obtained previously by TEM, we define
$g(D)$ as the distribution of number of particles having a
diameter equal to $D$. For spherical particles $\mu=\pi M_{sb} D^{3}/6$,
where $M_{sb}$ is, for the reasons given in the previous section,
taken as equal to the saturation magnetization of bulk Co. We fit the experimental
using the following expression:
\begin {equation}
M(H,T)=x_{para} \mu_{B} \tanh \left( \frac{\mu_{B}H}{k_{B}T} \right)+\\
(1-x_{para}) M_{sb} \frac{\int g(D) V M(H,T,\mu) dD} {\int g(D) V dD}
\label{Langevin}
\end{equation}
taking for $g$ a Gaussian distribution.

For each sample, we fit only data measured at temperatures for
which the two calculations, with and without anisotropy, give
approximately the same result. We give an example of this method
in the lower plot of Fig. 3 for a multilayer with $t_{Co}=0.3$ nm.
Above $30$ K, the calculations performed with and without
anisotropy almost coincide. For lower temperatures, close to
$T_{B}=8.6$ K, the experimental magnetization starts to deviate
from the pure isotropic behavior, as happens for the data measured
at $T=12$ K. Even then, the experimental data are rather well
reproduced by our calculations if we use the value of $K_{eff}$
determined from the blocking of the ac susceptibility (see section
\ref{SS.anisotropy} below). A list of the $\langle D \rangle$
values obtained from the fit for all samples is given in Table I.
We find that $\langle D \rangle$ increases with $t_{Co}$, as
expected. For most of the samples it is however larger by ten to
forty percent than the values that were previously found by TEM.
This discrepancy can be ascribed to the fact that the TEM
experiments were performed on single layers deposited on a special
carbon substrate, whereas we have measured multilayers prepared on
a Si oxide. However, we have also measured a monolayer with
$t_{Co}=0.7$ nm and obtained almost the same magnetization results
(see the upper plot of Fig. 3) as for a multilayer. An alternative
explanation is that the saturation magnetization of the smallest
particles is enhanced with respect to the bulk, as was found in
similar systems of Co clusters,\cite{Billas94,Chen95,Respaud98}.
However, even if we had used the maximum value of $2.3 \mu_{B}$
per Co atom, which was found by Respaud et al., $\langle D
\rangle$ would have decreased by only a $10$ percent, that is,
within the uncertainty of the fitting procedure. In order to get
the same diameters that were observed by TEM, $M_{s}$ should be as
large as $3 \mu_{B}$ to $4 \mu_{B}$ per Co atom for the smallest
clusters.

The fit of the magnetization curves is more sensitive to the value
of $\langle D \rangle$ than to the width of the distribution
$\sigma$. In fact, it is possible to obtain a reasonably good fit
by using a single Langevin curve with almost the same value of
$\langle D \rangle$ (see Fig. 3, upper plot). In order to get a
better estimation of $\sigma$, we have also fitted the equilibrium
susceptibility, as obtained from the low field dc or ac
measurements well above $T_{B}$, using the following expression
\begin{equation}
\chi_{eq}=\frac{\int g(D) V \left( M_{sb}^{2}V/3k_{B}T \right) dD}
{\int g(D) V dD}. \label{Eqsusc}
\end{equation}
This formula is valid also for particles with uniaxial magnetic
anisotropy if, as it is the case for our samples, the anisotropy
axes are not oriented.\cite{Shliomis94} It turns out that the
equilibrium susceptibility is very sensitive to the presence of
large particles in the distribution and therefore to $\sigma$, as
it is shown in the inset of Fig. 1. For this reason, the
contribution of the paramagnetic moments to $\chi_{eq}$ can be
neglected in all cases. The values for $\langle D \rangle$ and
$\sigma$ that are given in Table I are those which reproduce best
both the equilibrium magnetization isotherms and the equilibrium
susceptibility. The width of the distribution is found to be
rather constant and in good agreement with the value found
previously by TEM. The slight increase of $\sigma$ as the average
size of the aggregates decreases was also observed in the TEM
data.\cite{Maurice99} In conclusion, the co-deposition of Co and
Al$_{2}$O$_{3}$ gives us Co clusters of controllable size and with
a narrow and nearly constant distribution of diameters.

\subsection{Magnetic anisotropy}
\label{SS.anisotropy}
Next, we want to study the dependence of the effective anisotropy on
the cluster size. The anisotropy of the particles can be
estimated by comparing the average activation energy $\langle U
\rangle$ to the average volume $\pi/ 6 \langle D\rangle^{3}$. Usually
$\langle U \rangle$ is estimated as $\langle U \rangle = 25
k_{B}T_{B}$, where $T_{B}$ is the temperature of the maximum of
the ZFC susceptibility. However, this procedure leads to an
overestimation of the anisotropy because $T_{B}$ depends not only
on $\langle U \rangle$ but also increases with $\sigma$.\cite{Gittleman74,Luis99}

In order to get more reliable values we need a way to obtain the full distribution
of activation energies $f(U)$, and then to find which value of $U$
corresponds to particles with a diameter equal to $\langle D \rangle$.
Fortunately, $f(U)$ can be directly determined from ac susceptibility
data measured near the superparamagnetic blocking
temperature $T_{B}$.\cite{Luis99,Luis96} As mentioned above, the blocking
occurs when the average relaxation time becomes of the order of
$1/\omega$. It is therefore clear that the
temperature dependence of $\chi^{\prime}$ and $\chi^{\prime
\prime}$ near $T_{B}$ is determined by the distribution of $U$
among the particles. In order to relate $\chi^{\prime}$ and $\chi^{\prime
\prime}$ to the distribution $f(U)$ it is
common to assume that those particles having $U>U_{b}$
are fully blocked and the ones that fulfill the opposite condition
are in equilibrium. This hypothesis is reasonable because the relaxation time depends
exponentially on $U$ according to Eq. \ref{Arrhenius}.
For non-interacting particles, the relation that we were looking for reads as follows,
\begin {equation}
\chi^{\prime} \simeq \int_{0}^{U_{b}} \chi_{eq}(U,T)f(U)dU+
\frac{2}{3}\int_{U_{b}}^{\infty} \chi_{\bot }(U,T)f(U)dU
\label{su1}
\end {equation}

\begin {equation}
\chi^{\prime \prime} \simeq \frac{\pi}{2} k_{B}T \chi_{eq}\left(
T,U_{b}\right) f \left( U_{b}\right) , \label{su2}
\end {equation}
\noindent where $U_{b}=k_{B}T \ln \left(1/\omega\tau_{0} \right)$
is the activation energy of those particles having exactly
$\tau=t_{e}$ at a given temperature.
$\chi_{eq}=M_{sb}^{2}V/3k_{B}T$ and
$\chi_{\perp}=M_{sb}^{2}/2K_{eff}$ are the equilibrium
susceptibility and the reversible (high frequency limit)
susceptibility, respectively.\cite{Shliomis94} It follows from Eq.
\ref{su2} that $U f(U)$ can be directly determined by plotting
$\chi^{\prime \prime}$ versus the scaling variable $U_{b}$. In
Fig. 7 we show the result for a multilayer with $t_{Co}=0.7$ nm.
Similar results were obtained for the other samples.

It is important to note here that $f(U)$ is the fraction of the
total magnetic volume occupied by particles having the activation
energy $U$, since the susceptibility is mainly dominated by
the contribution of the largest particles. Contrary, $g(D)$ gives
instead the number of particles of a given size. The two distributions
are related as follows
\begin{equation}
f(U)=\frac{Vg(D)}{\left( dU/dD \right)
\int^{\infty}_{0}Vg(D)dD}\label{distributions}
\end{equation}
For spherical particles $U(D)=K_{eff}\left( \pi/6 \right) D^{3}$.
Therefore $f(U)$ is, apart from normalization factors,
proportional to $(U/K_{eff})^{1/3}g \left[ (6U/\pi K_{eff})^{1/3} \right]$.
Using this relationship and taking as above a gaussian $g(D)$,
it is possible to fit $\chi^{\prime \prime}(U_{b})$. The
anisotropy constant is then simply the ratio between $U(\langle
D \rangle)$ and $V( \langle D \rangle)$.
Although the fit is rather good, we find that the function $g$
that is extracted in this way from $f(U)$ (or $\chi ^{\prime
\prime}$) is systematically narrower than the size distribution obtained
previously using the equilibrium magnetization and magnetic susceptibility. We
will discuss later on the possible physical origin of this
discrepancy.

Before we comment on the variation of the anisotropy with the
size, we would like to show that the distribution $f(U)$ can also
be obtained by a different method, which makes use of the ZFC and
FC dc susceptibility curves measured at low enough magnetic
fields. The difference between the ZFC and FC magnetization curves
stems from the different contribution that the blocked particles
make to each of them. Neglecting the weak variation of $M_{sb}$
with $T$, this contribution only depends on $T$ via the critical
energy $U_{b}$ which determines the relative number of blocked and
superparamagnetic particles at a given temperature. Using the same
approximation which led to Eqs. \ref{su1} and \ref{su2} for the ac
susceptibility, it is possible to show that
\begin{equation}
\frac{\partial \left (M_{FC}-M_{ZFC} \right )} {\partial T} =
-M_{irr}\left( U_{b},T,H \right) f \left( U_{b} \right).
\label{ZFCFC}
\end{equation}
where $M_{irr}=M_{eq}-M_{rev}$ and $M_{rev}$ is the magnetization
brought by the reversible rotation of the magnetic moments. This
expression is valid provided that the applied magnetic field is
much smaller than the anisotropy field $H_{k}=2K_{eff}/M_{sb}$, as
it is actually the case in our experiments. If this condition was
not fulfilled, the activation energy would be a function of the
field and of its orientation with respect to the easy axes of the
particles. It is also possible to approximate $M_{irr}(U_{b})
\simeq \chi_{eq}(U_{b})H$. Therefore, Eq. \ref{ZFCFC} gives an
independent method to determine $f(U)$. We plot in Fig. 6 the
results obtained for an applied field of $10$ Oe, which are in
good agreement with the ac susceptibility data. In the same
figure, data obtained for $H=100$ Oe are also shown. In  this case
the maximum of the distribution shifts towards lower values of
$U_{b}$, indicating that the activation energy decreases in a
magnetic field. In addition, the distribution function broadens a
bit as a result of the random orientation of the easy axes.

We now come back to our main goal. The anisotropy constant is
plotted in Fig. 8 as a function of the average diameter of the
aggregates. It is interesting to compare these experimental data
with the constant $K_{eff}$ that is estimated using only the
intrinsic magnetocrystalline anisotropy of bulk Co. For hcp Co,
the stable phase for large particles, $K_{eff}$ equals the
intrinsic uniaxial anisotropy constant $K = 4.3 \times 10^{6}$
erg/cm$^{3}$. However, the structural characterization of all
samples studied here shows that they crystallize in the fcc phase.
Therefore, we would expect that the intrinsic anisotropy of the
particles in our samples would be smaller than for hcp Co. For
cubic anisotropy\cite{Gittleman74,Bean59} $K_{eff} = K/4$, where
$K$ is the second order intrinsic anisotropy constant. Taking $K =
2.8 \times 10^{6}$ erg/cm$^{3}$ for fcc Co, this gives $K_{eff} =
7 \times 10^{5}$ erg/cm$^{3}$. Therefore, the values that we find
for all samples are almost one to two orders of magnitude larger
than expected for magnetocrystalline anisotropy. Furthermore,
$K_{eff}$ is observed to increase as $\langle D \rangle$
decreases. The size dependence of the effective anisotropy follows
approximately the following phenomenological expression
\begin{equation}
K_{eff} = K_{\infty}+\frac{6K_{s}}{\langle D
\rangle}\label{anisotropy}
\end{equation}
with $K_{\infty} = 5 (2) \times 10^{5}$ erg/cm$^{3}$ and $K_{s} =
3.3(5) \times 10^{-1}$ erg/cm$^{2}$. This result is robust in the sense
that it does not change qualitatively if we use the average
diameter found by TEM, instead of the values obtained from the
magnetization data. The first term is close to $K/4$ and can
therefore be identified as the contribution of
the intrinsic anisotropy. The second one is proportional to the
fraction of atoms located at the periphery of the particles, which
can be more than $80$ \% of all Co atoms for the smallest clusters studied here. Our
experimental results indicate then that there exists a rather
large contribution of the surface of the particles to the net
anisotropy.

The enhancement of the magnetic anisotropy of nanometer sized
metallic particles with respect to the bulk has been previously
reported by several authors \cite{Bodker94,Chen95,Respaud98}. For
Co particles with diameters varying between $4.4$ and $1.8$ nm,
Chen and coworkers\cite{Chen95} obtained $K_{eff}$ which increases from about $5
\times 10^{6}$ erg/cm$^{3}$ to about $3 \times 10^{7}$
erg/cm$^{3}$. These values are even larger than ours.
However, they are of the same order as the values that would have
been obtained if we had used the temperature of the maximum of the ZFC
susceptibility, as it was done by the authors. More recently,
Respaud et al.\cite{Respaud98} studied the anisotropy of Co particles of $1.5$ and
$1.9$ nm by fitting the whole ZFC and FC magnetization curves, a
method that can be considered as equivalent to ours. They found
$K_{eff} \simeq 8.3 \times 10^{6}$ erg/cm$^{3}$ and $K_{eff}
\simeq 7.3 \times 10^{6}$ erg/cm$^{3}$, respectively, in
reasonably good agreement with our data. The existence of a large
surface anisotropy in metallic particles is thus well established
experimentally.

The origin of this extra anisotropy has been related to the
modification of the electrostatic and exchange interactions of the
atoms located at the surface,\cite{Pastor95,Bruno89,Kodama99}
which depends largely on whether the surface is oxidized or not.
However, as we argued above, the characterization of our samples
by EXAFS, EELS, and XPS does not indicate the presence of an oxide
layer.\cite{Briatico99} The same conclusion is derived from the
hysteresis loops measured below $T_{B}$. Therefore, we have to
consider how the properties of a "bare" metallic surface are
modified with respect to the bulk. The value of $K_{s}$ that we
have found is actually comparable to the perpendicular anisotropy
measured in free Co surfaces.\cite{Weller95} It is commonly
accepted that this perpendicular anisotropy is related to the
appearance of a large orbital magnetic moment on these
atoms.\cite{Wang93} The $3$d electrons become more localized at
the surface and the localization gives rise to an increase of the
orbital moment. The same theoretical interpretation can be applied
to the atoms at the periphery of small metallic
clusters.\cite{Pastor95} In this case, the enhanced anisotropy at
the surface extends to the inner atoms via the strong exchange
interaction with them, which leads to an increase of the average
anisotropy of even spherical clusters.\cite{Dimitrov94} This
interpretation has been confirmed by X-ray magnetic dichroism
experiments performed on Au/Co/Au layers\cite{Weller95} and more
recently also on Co disk-like aggregates supported on Au
surfaces.\cite{Durr99}. It was found that the orbital component
$m_{L}$ of the total magnetic moment scales with the fraction of
atoms located at the surface of the aggregates. For spherical
clusters, as the ones studied here, we expect then that $m_{L}
\propto 1/\langle D \rangle$, dependence that we have indeed
observed for $K_{eff}$. We therefore conclude that the observed
increase of $K_{eff}$ is likely due to the increasingly localized
character of the $3$d electrons of the atoms located at the
surface.

Once the particle size distribution and the anisotropy are known,
it is possible to predict the time-dependent magnetic response of
the samples and compare it to the experiment. Examples of these
calculations are compared to the experiments in Figs. 1, 3 and 5.
The calculations account very well for the experimental data
measured above $T_{B}$, as expected. They also reproduce in Fig. 1
the deviation of the FC susceptibility from the equilibrium
susceptibility that takes place below $5$ K. However, they
reproduce neither the position nor the shape of the maximum of the
ZFC susceptibility. Another example of this discrepancy is shown
in Fig. 9, where we plot the experimental $\chi^{\prime}$ for a
multilayer with $t_{Co} = 0.3$ nm and the values calculated
(dotted line) with Eq. \ref{su1}. Again, the width of the blocking
transition is clearly overestimated by the calculations.

We recall here that we have found that the activation barriers
distribution is systematically narrower than the size distribution
for all samples. As an example, in Fig. 10 we plot the size
distribution $g(D)$ of a multilayer with $t_{Co} = 0.7$ nm
extracted from $\chi^{\prime \prime}$ and directly observed by
TEM. The horizontal scale for the former distribution is
$(6U_{b}/\pi K_{eff})^{1/3}$, with $K_{eff} = 10^{7}$
erg/cm$^{3}$. It is tempting now to attribute the "narrowing" of
the blocking transition to the effect of the surface anisotropy.
When $K_{s}/D \gg K_{\infty}$ then $U \approx K_{s} S$, where
$S=\pi D^2$ is the surface of the particle. It follows then from
Eq. \ref{distributions} that $f(U) \propto D^{2} g(D)$ and the
width of the distribution of activation energies must then be
smaller than when $U \propto V$. Figure 10 shows indeed that when
the same susceptibility data are represented versus the variable
$(U_{b}/\pi K_{s})^{1/2}$ the ensuing size distribution is in
better agreement with what it is found by TEM or from the
equilibrium magnetization and susceptibility. In this way, we also
obtain $K_{s}$ which turns out to be between $0.2$ and $0.3$
erg/cm$^{2}$ for all samples. This value can be then used to
recalculate the ac susceptibility and the ZFC magnetization. We
find that the calculations performed with the same parameters
$\sigma$ and $\langle D \rangle$ as before (see Table I) but
taking $U \propto D^{2}$ are in much better agreement with the
experiment (see Figs. 1 and 9). Although the width of the of size
distribution is not always accurately determined, it seems that
the influence of the surface anisotropy also modifies the shape of
the susceptibility peak at the blocking. We conclude that the
dynamical response of very small particles is therefore determined
by the special physical properties of the atoms which are located
at their surface.

\subsection{Influence of the number of layers: dipole-dipole
interaction between the particles} \label{SS.inter}

There has been some debate during the last years about the effect
that the dipole-dipole interaction between magnetic nano-particles
has on their relaxation times.  Shtrikman and Wolfarth\cite{Shtrikman81} and later
Dormann et al.\cite{Dormann88} predicted that the effective
activation energy increases by an amount that depends on the
number and spatial arrangement of the neighbor particles. By
contrast, in the model proposed by S. M{\o}rup and E. Tronc
\cite{Morup94} the interaction between the particles leads to a
lower $U$. The experimental validation of one of these two models
is complicated because, for some preparation methods, it is
difficult to vary the density of particles in the sample without
modifying the distribution of particle's sizes.\cite{Dormann88} A different
approach is to dissolve the particles in a fluid and to change the
concentration by varying the amount of solvent. However, it is
possible that the particles agglomerate in the fluid because
of their mutual interaction, so that the interaction with the
nearest neighbors is not greatly affected.\cite{Chantrell96}

The preparation method of our samples presents a number of
advantages. We have seen that the average size can be controlled
by changing the deposition time, but also the packing of the
particles can be controlled. The TEM images show that the clusters
in a layer do not agglomerate and, furthermore that the deposition
of several layers of Co and Al$_{2}$O$_{3}$ leads to a
self-organized spatial arrangement of the particles (see ref.
\cite{Babonneau00}). For a multilayer each cluster has, in
average, six nearest neighbors in the same plane, three above and
another three below it. For $t_{Co}=0.7$ nm, the average distance
between nearest Co clusters in the same layer is $\Lambda_{\|}
\simeq 5.4$ nm, whereas the distance to nearest neighbors in
adjacent layers is $\lambda = 4.5$ nm.\cite{Babonneau00} In this
section, we compare the relaxation rate of two samples having both
$t_{Co}=0.7$ nm ($\langle D \rangle \simeq 3$ nm), but very
different number of layers, namely $30$ and only one. By going
from a monolayer to a multilayer we certainly expect that the
average energy of interaction of a particle with the others
changes. The interaction energy between particles in adjacent
layers is the largest and of the order of $\mu^{2}/\lambda^{3}
\approx 40$ K. By contrast, in a sample with a single layer, each
particles has, in average, only six neighbors coupled by a weaker
interaction ($\mu^{2}/\Lambda_{\|}^{3} \approx 20$ K).

In order to attribute any difference between the two samples to
the effect of the interparticle interactions, it is very important
to check beforehand that the sizes of the aggregates are the same
in both. We showed in Fig. 3 that the equilibrium magnetization
curves of the two samples are almost identical, and we compare in
Fig. 11 the inverse of their ac susceptibility curves. Above
$T_{B}$, the susceptibility follows the Curie-Weiss law, with
identical values of $C$, which confirms that $\langle D \rangle$
and $\sigma$ are practically the same. By contrast, the
Curie-Weiss temperature $\theta$ is about $2$ times smaller for
the monolayer, indicating that the average inter-particle
interaction is notably reduced. It is also apparent that the
blocking temperature of the monolayer is smaller than that of the
multilayer. As we have done before, the activation energy of the
two samples can be compared by plotting $\chi^{\prime \prime}$
measured at different frequencies as a function of the scaling
variable $U_{b}$. This comparison is shown in Fig. 12. The maximum
of the curve for the monolayer is clearly shifted towards lower
values of $U_{b}$ with respect to the maximum obtained for the
multilayer. Our data give strong evidence that the interaction
between the aggregate layers tends to increase the activation
energy of each particle, by an amount of about $200$ K. This
difference is of the same order of magnitude as the interaction
energy with the six nearest neighbors in the multilayer. We also
find that the relative width of $U f(U)$ has the same value for
the two samples, which confirms again that the distribution of
particle's sizes is the same.

\subsection{Magnetic relaxation at low temperatures} \label{SS.rel}

In the previous sections, the reversal of the magnetic moments has
been treated as a classical process assisted by the interaction
with a thermal bath. However, taken as a quantum variable, the
spin of a magnetic particle $S=M_{s}V/g\mu_{B}$ can in principle
flip also by quantum tunneling across the barrier if the effective
Hamiltonian contains terms which deviate from the uniaxial
symmetry.\cite{Chudnovsky98} This possibility is very attractive because it would
show the existence of quantum effects at the intermediate scale
between the microscopic and the macroscopic worlds. Quantum
relaxation can dominate over the thermal activation at very low
temperatures, when the thermal population of the first excited
state doublet $\pm (S-1)$ becomes negligible, and should lead to
a saturation of the relaxation rate to a nearly temperature
independent value.\cite{Prokofev96} Such a saturation has indeed been observed in
some systems of single domain particles in the
past.\cite{Gunther95,Chudnovsky98,Peng99,Balcells92}

In this section, we present measurements of the relaxation of the
remanent magnetization of an initially saturated sample. We have
chosen the sample with the smallest Co clusters for two reasons;
first, because the rate for quantum relaxation must be the largest
for these clusters of only about $50-100$ atoms; and second,
because this sample shows the strongest anisotropy. The separation
of the two lowest lying state doublets, which is roughly given by
$\Omega_{0} \approx g\mu_{B}H_{k}$ is then about $3$ K, thus
larger than the lowest temperature that our magnetometer can reach
($T_{min} = 1.7$ K). We have measured the decay of the
magnetization of the sample that takes place after a magnetic
field of $5$ T is switched off at different temperatures. The
decay of $M_{r}$ is approximately logarithmic in time. An
important advantage of recording the relaxation at zero field is
that it can then be easily calculated using our knowledge of the
activation energies distribution. At zero field, the equilibrium
magnetization is zero for all particles. Therefore, using the same
approximation as before, the time dependent magnetization is given
by
\begin {equation}
M(t,T)=\frac{M_{s}}{2} \int_{U_{b}}^{\infty}
f(U)d(U)\label{relaxation}
\end{equation}
where we have made the reasonable approximation that the magnetic
moments of the particles are initially saturated by the magnetic
field. The factor $1/2$ arises from the reversible rotation of the
magnetic moments for a random orientation of the easy axes, as in
the Stoner-Wolhfarth model.\cite{Stoner48} As pointed out by
Labarta et al.,\cite{Labarta93} if the magnetic moments flip by a
thermally activated process the relaxation curves measured at
different temperatures should scale when plotted as a function of
$U_{b}$. This plot also gives a picture of the relaxation at very
long times, which are not experimentally accessible. Our
experimental data, which we plot in Fig. 13 do indeed show a
rather good scaling for the same $\tau_{0}=10^{-13}$ s that was
obtained from the shift of the maximum of $\chi^{\prime \prime}$
with frequency. The full line in the same figure was calculated
with Eq. \ref{relaxation} using the distribution $f(U)$ that we
determined with the method described in section
\ref{SS.anisotropy}. The scaling of the data confirms that the
relaxation mechanism is classical (not tunneling) down to $T=1.7$
K.

In the inset of the same Fig. 13 we show the temperature
dependence of the so-called magnetic viscosity $S_{r}$, determined
as the slope of the $M_{r}$ vs $\ln(t)$ curves. Below about $2.5$
K, $S_{r}$ does not vary much with $T$. We note however that,
according to Eq. \ref{relaxation}, the magnetic viscosity is just
\begin {equation}
S_{r} \equiv \frac{\partial M}{\partial \ln t}=-k_{B}T
\frac{M_{s}}{2}f \left( U_{b} \right)\label{viscosity}
\end{equation}
and it is therefore proportional to $f(U)$. The apparent
saturation of $S$ measured between $1.7$ K and $2.5$ K just
reflects the shape of the distribution $f(U)$, and it is indeed
rather well described by the "classical" calculation. These data
give an example of how important it is to have information about
$f(U)$ in order to adequately interpret the relaxation
data.\cite{Barbara93}

\section{FINAL REMARKS AND CONCLUSIONS}
\label{SDis} We have presented a detailed and extensive study of
the magnetic properties of Co aggregates prepared by sequential
deposition of Co and Al$_{2}$O$_{3}$. This preparation method
enables us to control both the average size and the number of
layers independently. We have shown that the distribution of
activation energies can be accurately determined from ac
susceptibility and ZFC-FC magnetization measurements. We have
investigated the variation of the effective anisotropy as the size
of the aggregates decreases from about $5$ nm to below $1$ nm. We
find that $K_{eff}$ scales with the fraction of atoms located at
the periphery of the aggregates. The strength of the surface
anisotropy is of the same order of what is found for free Co
surfaces and we therefore attribute it to the increase of the
orbital magnetic moment of these atoms. Furthermore, the
activation energies distribution resembles the distribution of
particle's surfaces rather than the volume distribution. For these
small clusters, it is therefore more appropriate to write
$U=K_{s}S$ than the "traditional" $U=K_{eff}V$. Using the
distributions of sizes and of activation energies that we have
determined, we are able to give a quantitative account of all the
equilibrium and time-dependent experimental quantities. We have
also shown that the activation energy increases when the average
number of nearest neighbors per particle increases, in agreement
with the model of Dormann et al. Finally, the decay of the
remanent magnetization of clusters containing only about $50$ to
$100$ atoms is shown to proceed via a thermally activated
mechanism down to the lowest temperatures investigated.

\textbf{ACKNOWLEDGMENTS} We would like to thank Dr C. Paulsen and
Dr. J. Carrey for assistance with some of the experiments reported
in this work. This work has been partly funded by Spanish Grant
MAT 99/1142 and the European ESPRIT contract "MASSDOTS".

\noindent $^*$ To whom all correspondence should be addressed.
E-mail address: barto@posta.unizar.es


\newpage

\begin{table}
\begin{tabular}{c c c c c c}
             & \multicolumn{2}{c}{(a)} & \multicolumn{2}{c} {(b)} & \\
$t_{Co}$ (nm) & $\langle D \rangle^{a}$ (nm) & $\sigma^{a}$ &
$\langle D \rangle^{b}$ (nm) & $\sigma^{b}$ & $x_{para}$\\ 0.1  &
& & 0.8(1) & 0.35(5)  & 0.7\\ 0.2  & 0.83(20)  & 0.3  & 1.3(1) &
0.3(1) & 0.7\\ 0.3  & 1.4(3)  & 0.3 & 1.4(1) & 0.32(5)  & 0.22\\
0.4 & 1.4(3) & 0.22 & 2.2(1) & 0.2(1)   & 0.24\\ 0.7  & 2.9(6) &
0.23 & 3.1(3) & 0.2(1)   & 0.25\\ 1    & 4.2(8)   & 0.27 & 5.2(3)
& 0.25(5) & 0.13\\
\end{tabular}
\caption{ Parameters of the gaussian distribution of particle's
sizes obtained by TEM (a) and from the fit of the magnetization
data (b). The width $\sigma$ of the distribution is given in units
of the average diameter. The last column gives the estimated
fraction of Co atoms which do not aggregate in particles.} \label{TabI}
\end{table}

\newpage


\settowidth{\columnwidth}{aaaaaaaaaaaaaaaaaaaaaaaaaaaaa
aaaaaaaaaaaaaaaaaaaa}

\textbf{Figure captions}

{\bf Figure 1}. dc susceptibility of a multilayer with
$t_{Co}=0.1$ nm and $N=100$ measured with a field of $100$ Oe;
open symbols, FC; closed symbols, ZFC. The lines represent the
results of calculations performed with the parameters of the size
distribution given in Table \ref{TabI}: dashed line, equilibrium
susceptibility; dotted line, ZFC susceptibility calculated taking
$U=K_{eff}V$ and $K_{eff} = 2.4 \times 10^{7}$ erg/cm$^{3}$; full
lines, ZFC and FC susceptibilities calculated for surface
anisotropy with $K_{s} = 0.3$ erg/cm$^{2}$. {\bf Inset}: Inverse
suceptibility of three multilayers: (a), $t_{Co} = 0.1$ nm,
$N=100$; (b), $t_{Co} = 0.3$ nm, $N=40$; (c), $t_{Co} = 0.7$ nm,
$N=30$. The lines represent the equilibrium susceptibility of (b)
calculated for $\langle D \rangle=1.4$ nm and three values of
$\sigma$: $0.25$ (upper curve), $0.3$ (medium curve), and $0.35$
(lower curve).


{\bf Figure 2}. Real and imaginary parts of the ac susceptibility
of a multilayer with $t_{Co}=0.3$ nm and $N=40$.

{\bf Figure 3}. Equilibrium magnetization of multilayers with
$t_{Co}=0.7$ nm (a) and $t_{Co}=0.3$ nm (b), measured at different
temperatures. The lines represent the calculated results. (a):
dotted line, pure Langevin curve for $D=3.1$ nm; full lines,
results calculated averaging a Langevin curve over a Gaussian
distribution of sizes (see Eq. \ref{Langevin}) with $\sigma=0.2$
and three different values of the average diameter. (b): full
line, as in the upper picture for $\sigma=0.32$ and $\langle D
\rangle = 1.4$ nm; dotted lines, equilibrium magnetization
calculated for $T=12$ K and $T=30$ K with the same size
distribution but for uniaxial anisotropy with $U=\pi K_{s}D^{2}$
and $K_{s} = 0.2$ erg/cm$^{2}$.

{\bf Figure 4}. Hysteresis loop of a multilayer with $t_{Co}=0.1$
nm measured at $T=2$ K after cooling the sample in zero field or
in $5$ T from room temperature.

{\bf Figure 5}. {\bf Left axis}: Temperature dependence of the
remanent magnetization and of the saturation magnetization
(measured with $H=50$ kOe) of a multilayer. The lines are
calculated with Eqs. \ref{Langevin} and \ref{relaxation},
respectively using the parameters given in Table I and the
distribution $f(U)$ estimated from the blocking of the ac
susceptibility. {\bf Right axis}: Temperature dependence of the
coercive field of the same sample.

{\bf Figure 6}. Variation of the reduced remanent magnetization
$m_{r}$ (open symbols, right axis) and of the low temperature
saturation magnetization $M_{s}$ (closed symbols, left axis) with
the amount of deposited Co for all samples studied.

{\bf Figure 7}. Imaginary part of the susceptibility of a
multilayer with $t_{Co}=0.7$ nm and $N=30$ plotted as a function
of the scaling variable $U_{B}/k_{B}=T \ln \left(1/\omega \tau_{0}
\right) $, with $\tau_{0}=10^{-13}$ seconds. The full line is a
fit according to Eq. \ref{su2} taking a Gaussian for $g(D)$.
Results obtained as explained in the text (cf Eq. \ref{ZFCFC})
from ZFC-FC magnetization curves measured with two different
magnetic fields are also shown for comparison.

{\bf Figure 8}. Size-dependence of the effective anisotropy
constant for all samples investigated. The full line is a best
squares fit of the data to Eq. \ref{anisotropy}.

{\bf Figure 9}. Real part of the susceptibility of a Co multilayer
with $t_{Co}=0.3$ nm and $N=40$ measured for two different
frequencies. The dotted line is calculated for $\omega/2 \pi =0.1$
Hz with Eq. \ref{su1} using the parameters of Table \ref{TabI} and
taking $U=\pi K_{eff} D^{3}/6$, with $K_{eff} = 1.15 \times
10^{7}$ erg/cm$^{3}$. The full lines are calculated taking $U =
\pi K_{s}D^{2}$, with $K_{s} = 2 \times 10^{-1}$ erg/cm$^{2}$.

{\bf Figure 10}. The size distribution determined by TEM is
compared to the distributions obtained from $\chi^{\prime \prime}$
for two limiting cases where the anisotropy is either dominated by
the intrinsic (volume) contribution (full line) or by the surface
anisotropy (dotted line). The scaling in the horizontal axis gives
respectively $K_{eff}=10^{7}$ erg/cm$^{3}$ and $K_{s}=0.33$
erg/cm$^{2}$.

{\bf Figure 11}. Inverse ac susceptibility of two samples with the
same $t_{Co}=0.7$ nm but different number of layers: open symbols,
$N=30$; full symbols, $N=1$.

{\bf Figure 12}. Scaling plot of $\chi^{\prime \prime}$ for two
samples with $t_{Co}=0.7$ nm but different number of layers. For
both $\tau_{0}=10^{-13}$ seconds.

{\bf Figure 13}. Time-dependent remanent magnetization of a (Co
$0.1$ nm Al$_{2}$O$_{3}$ $3$ nm)$_{100}$ multilayer plotted as a
function of the scaling variable $U_{B}/k_{B}=T \ln \left(t_{e}/
\tau_{0} \right)$ with $\tau_{0}=10^{-13}$ seconds. The inset
shows the temperature dependence of the magnetic viscosity. The
full lines are calculated according to Eqs. \ref{relaxation} and
\ref{viscosity}.


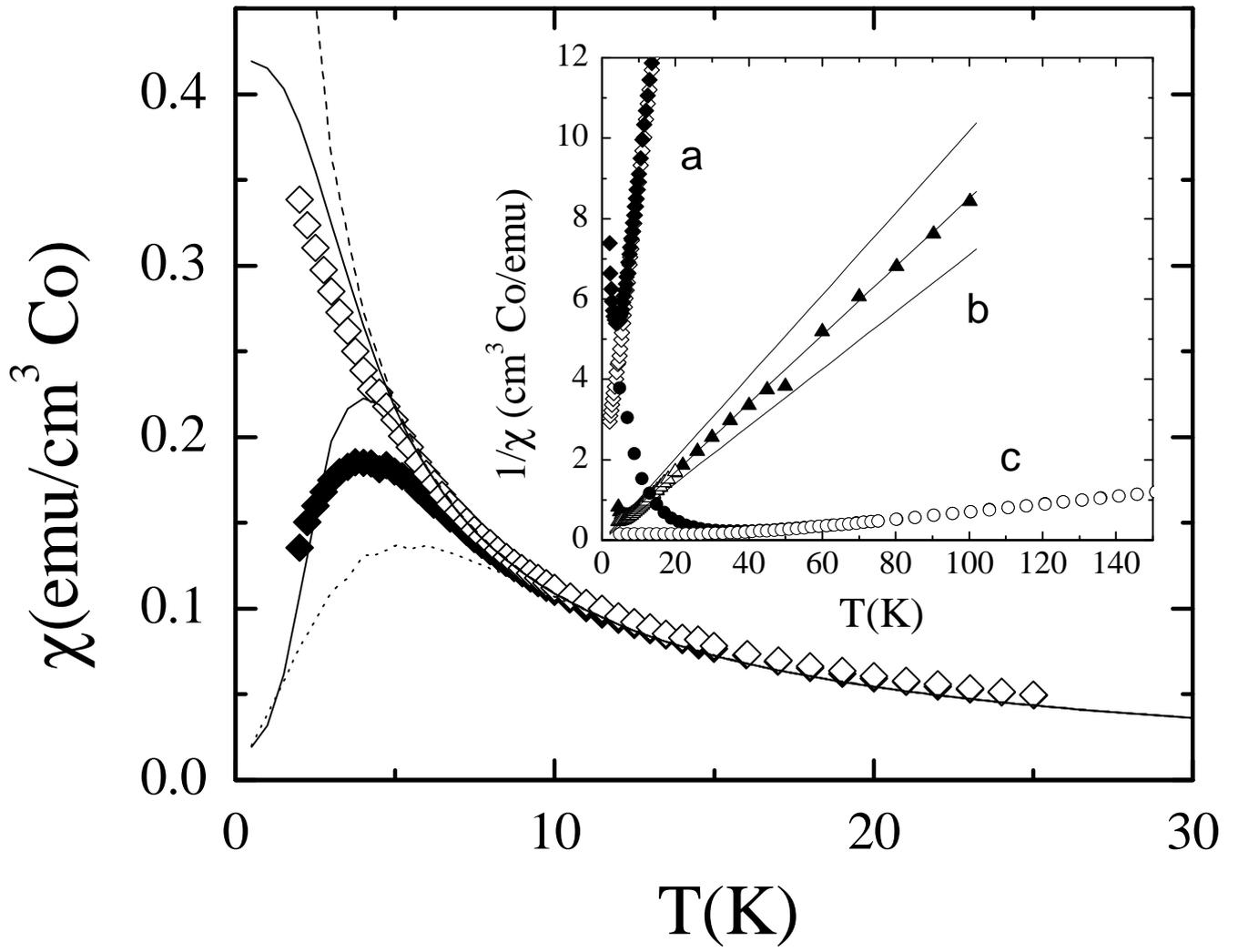
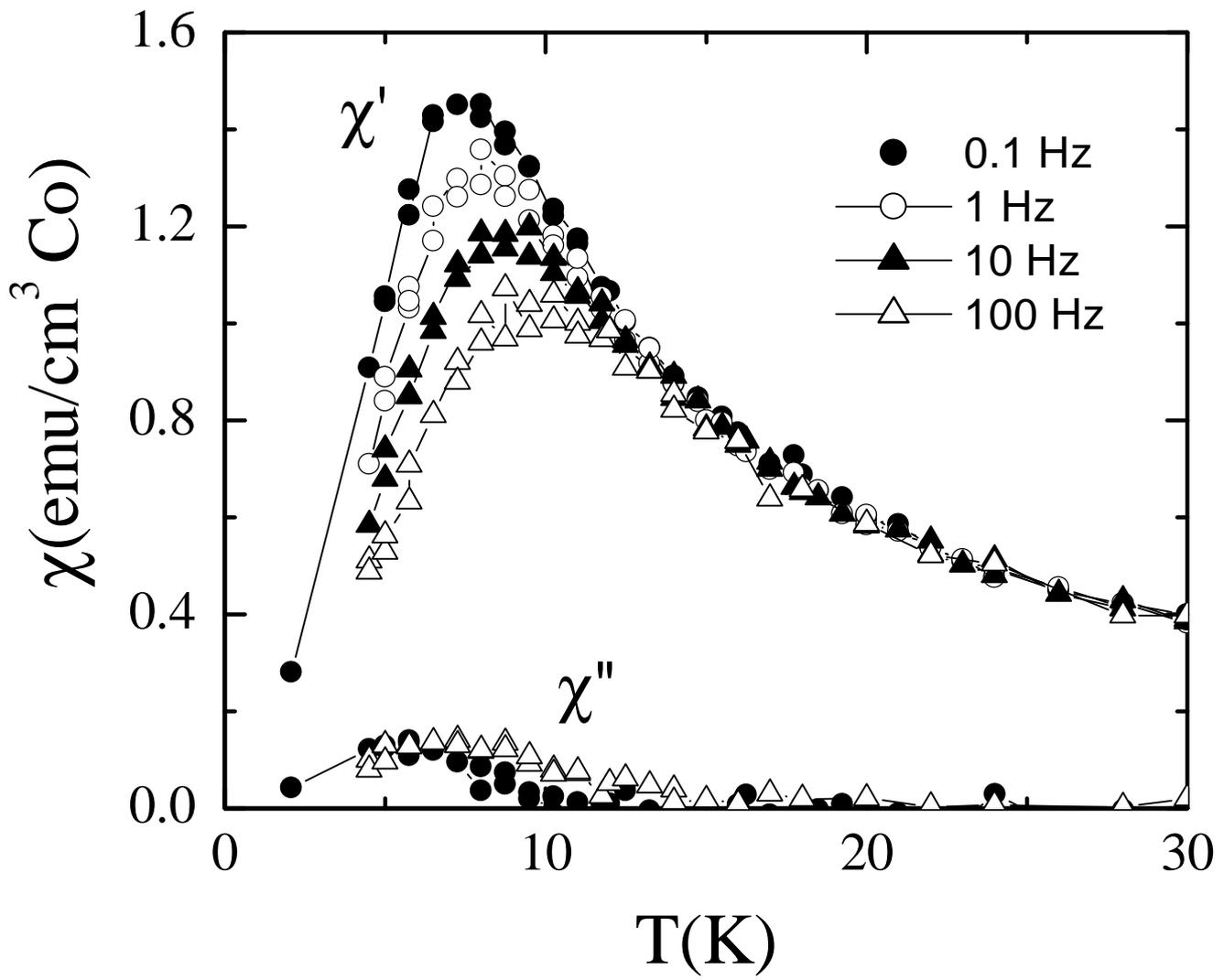
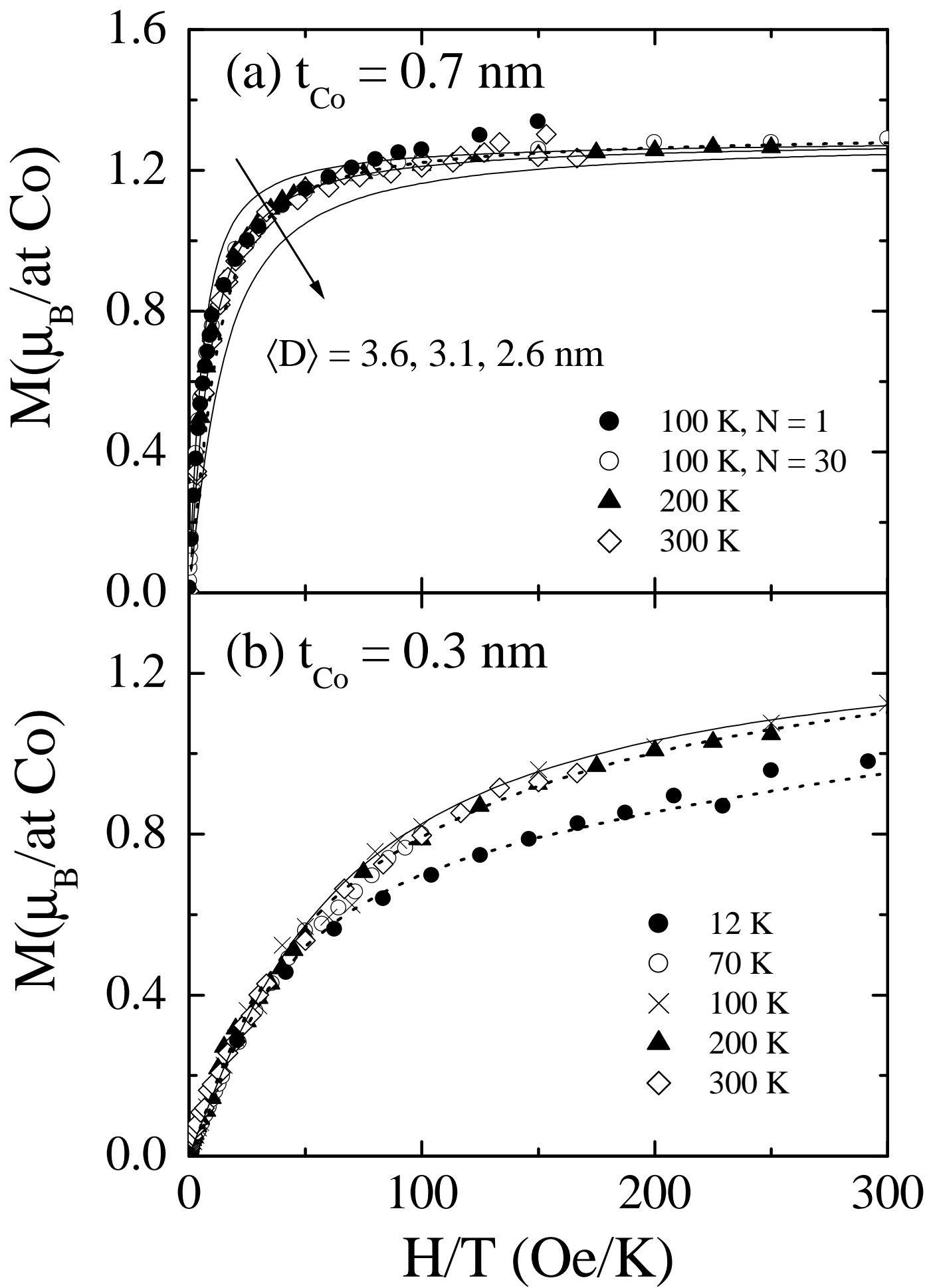
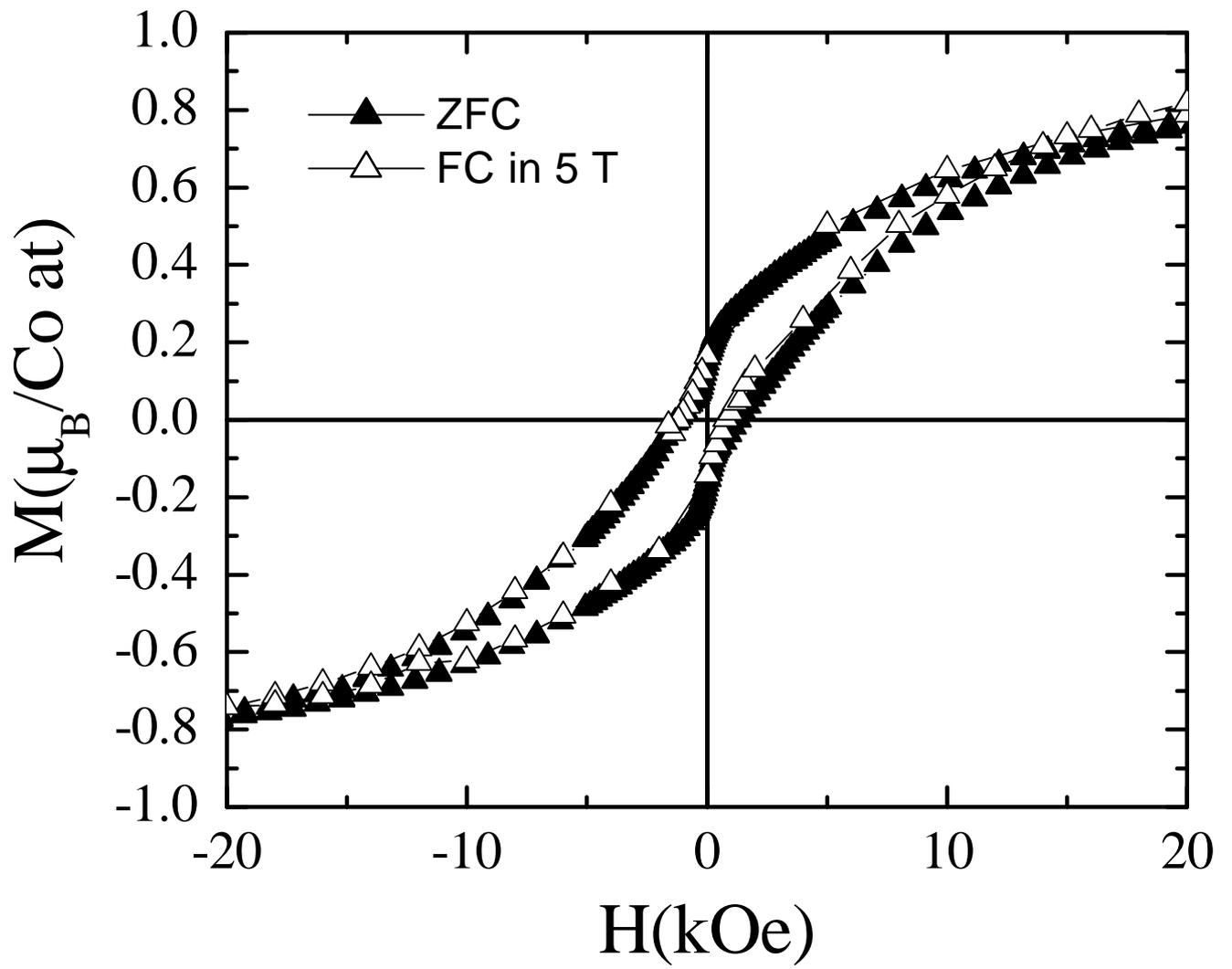
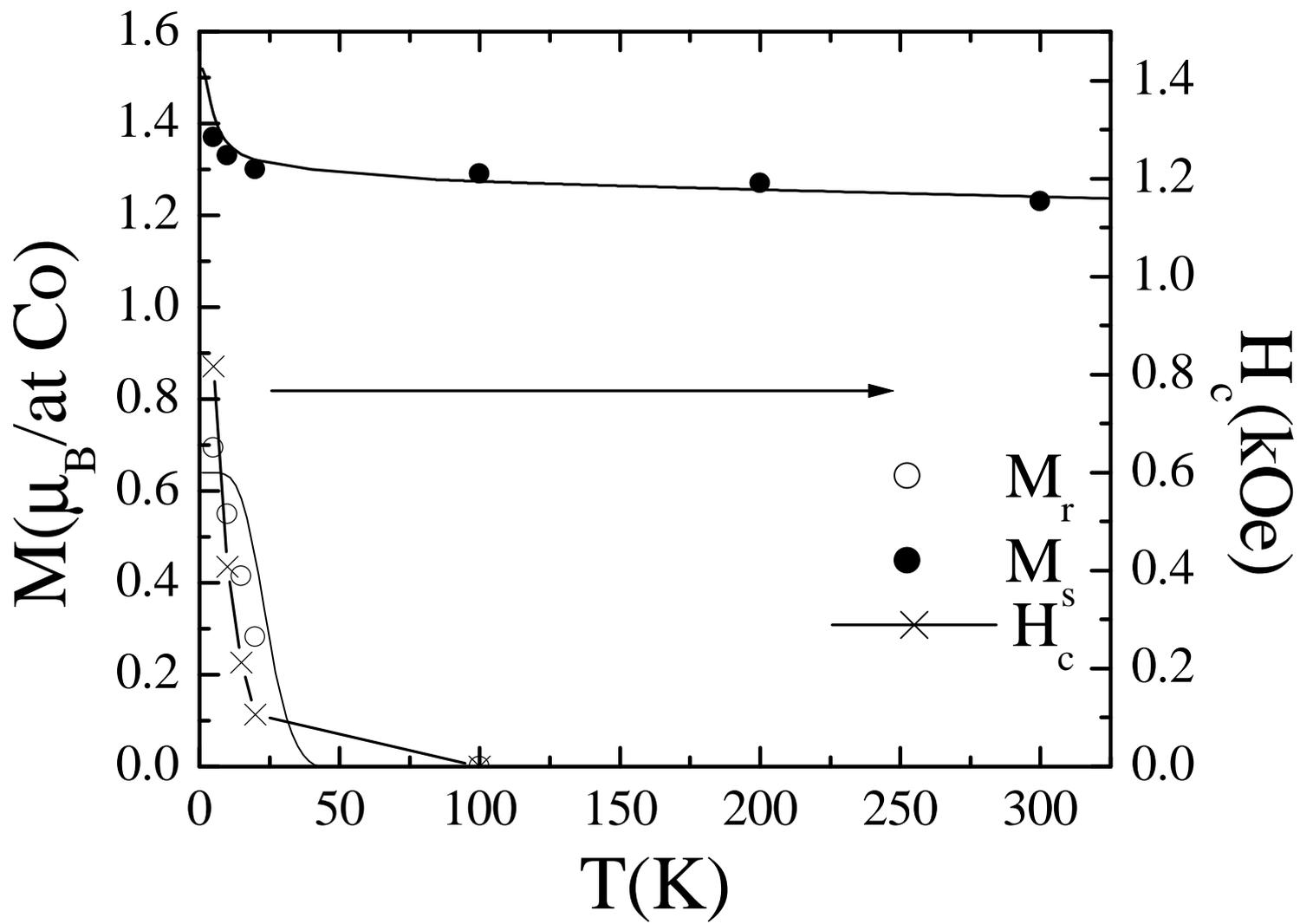
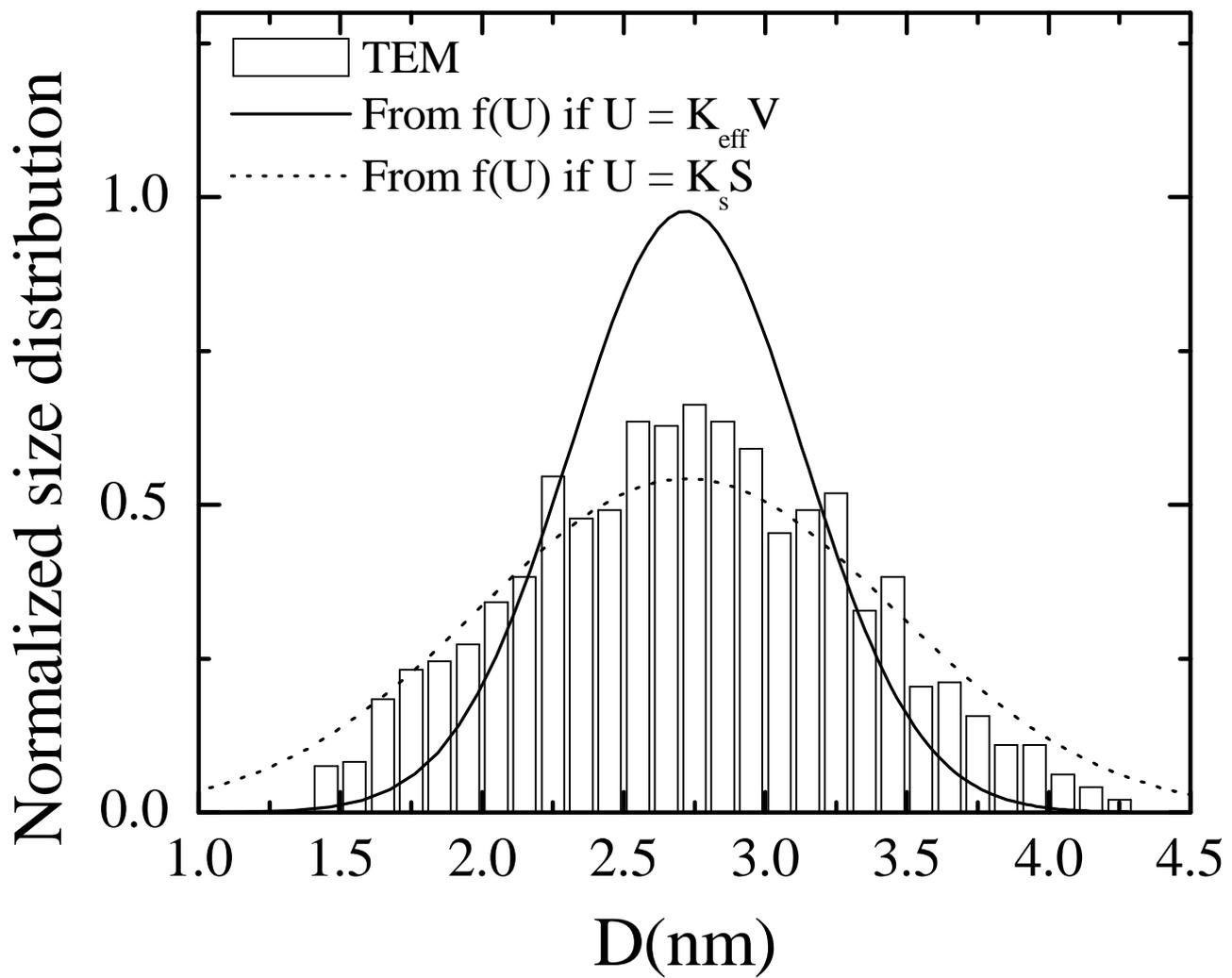
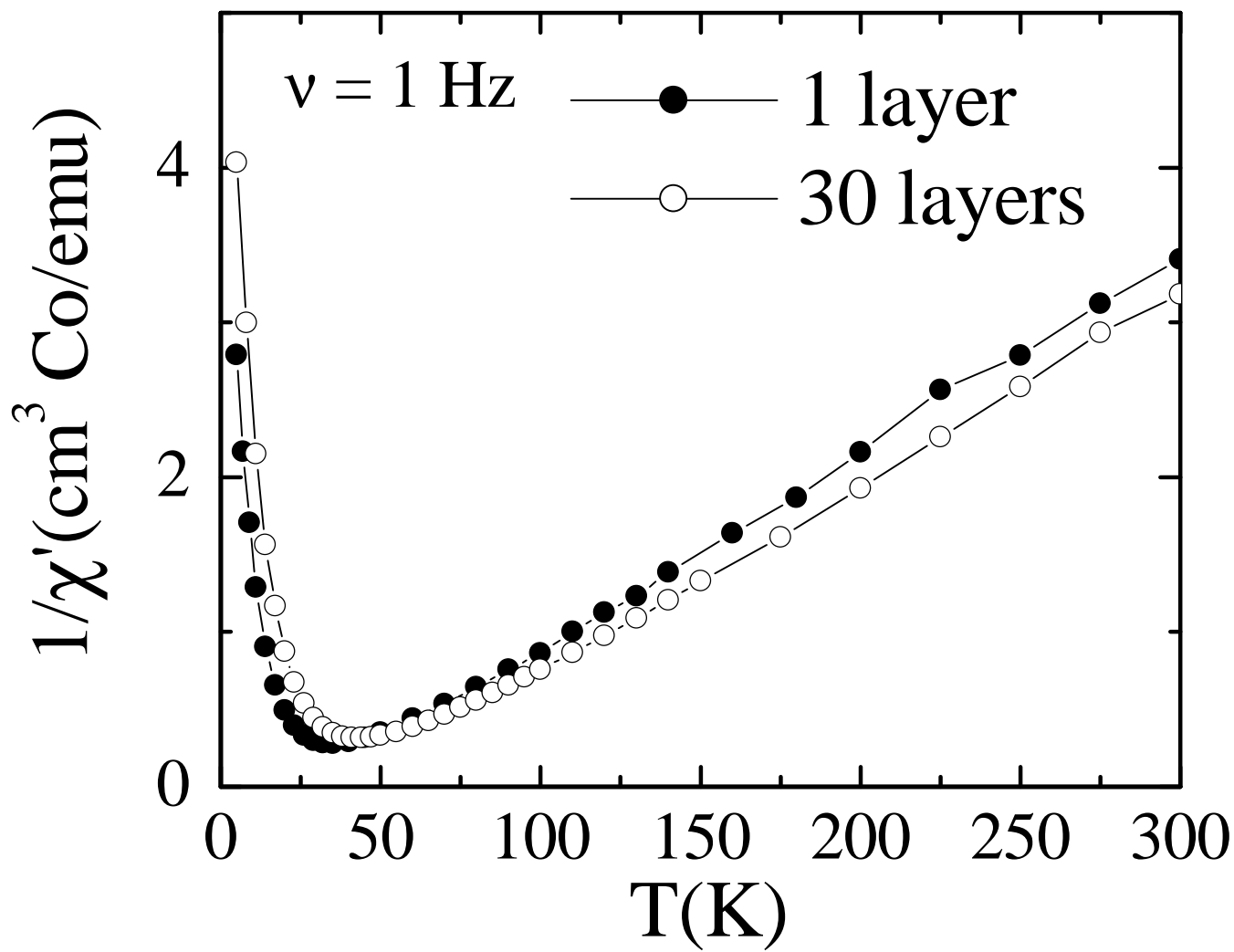
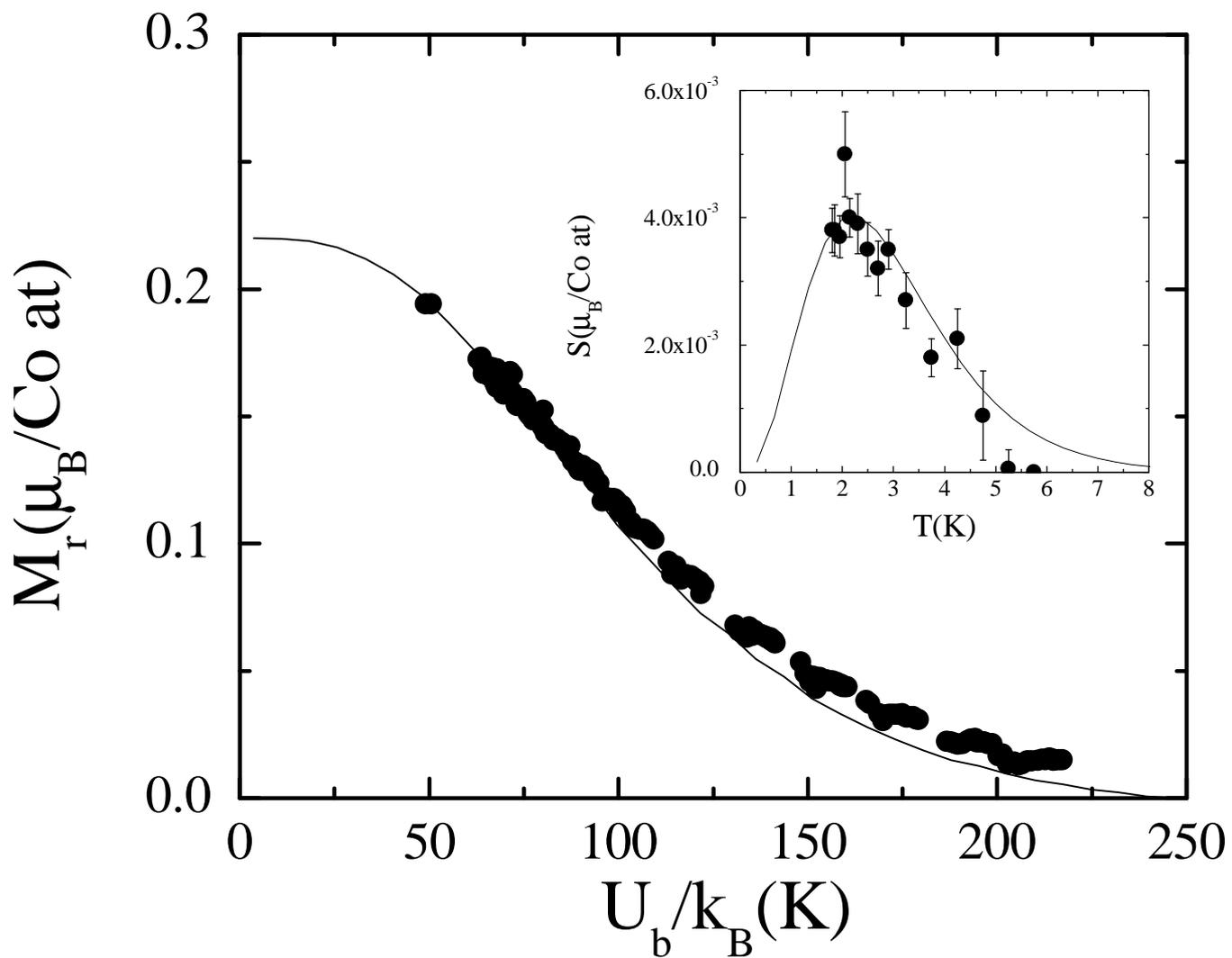
\end{document}